\documentclass[%
 reprint,
 amsmath,amssymb,
 aps,
]{revtex4-2}

\usepackage{graphicx}
\usepackage{dcolumn}
\usepackage{bm}
\usepackage{physics}
\usepackage{hyperref}
\usepackage{mathrsfs}

\begin{document}

\title{Quantum Nature of Quasi-Classical States and Highest Possible Single-Photon Rate}

\author{Moslem Mahdavifar}
\email{mahdavifar.m@miami.edu}
\affiliation{Department of Physics, University of Miami, Coral Gables, Florida 33146, USA}

\renewcommand{\figurename}{Fig.}

\def\pmid{p_{mid}}
\def\SIn{\sigma_\pm^{in}}
\def\SOut{\sigma_\pm^{out}}
\def\SD{D_{\pm}}
\def\SH{h_{\pm}(x,t)}

\begin{abstract}
Observation of the purely quantum mechanical effects of quasi-classical states is of utmost importance since these states are realistic sources of radiation and do not have any shortage in photon numbers. Therefore, they do not face the scalability problem as much as other single-photon sources do, which makes them much more robust against photon loss. Moreover, these states define the standard quantum limit. Hence, finding their quantum signature hints to the highest possible single-photon rate. In this manuscript, we attempt to demonstrate this idea theoretically using known dynamics and then present supporting experimental results. Through our experiment, we realize two-photon bunching from the transfer of quantum information using such states with the projection of orbital angular momentum from a continuous wave source. Our work is a step forward towards a more diverse and practical use of quasi-classical states in the domain of quantum optics and quantum information.
\end{abstract}

\maketitle

Observation of two-photon bunching and its accompanying Hong-Ou-Mandel (HOM) effect in an interferometeric setting is the signature of the quantum nature of light, and upon this observation we need a fully quantized frame to correctly describe the radiation \cite{hong1987measurement}. Sources satisfying this criterion can be safely used to probe the quantum world. This non-classical behavior is at the core of many modern quantum applications such as quantum teleportation, quantum computation and quantum metrology \cite{bouwmeester1997experimental,knill2001scheme,giovannetti2011advances}. Purely quantum mechanical sources such as heralded photons from Spontaneous Parametric Down Conversion (SPDC) suffer from low photon numbers due to poor efficiency of non-linear processes that produce them \cite{boyd2020nonlinear,bachor2019guide,eisaman2011invited}. This is an especially acute limitation if further optical processing is expected. In this regard, it has been highly desirable to find alternative sources that can overcome this shortage. Radiation from a continuous wave source will be described by a set of states known as quasi-classical or coherent states. The quasi-classical states are quantum states with interesting characteristics. These states are the eigenstates of the annihilation operator and like the vacuum state, they define the standard quantum limit (SQL), i.e., they minimize the uncertainty relations \cite{sudarshan1963equivalence,glauber1963coherent}. In addition, they predict correct behavior of physical quantities like the average electric field in the classical regime \cite{grynberg2010introduction,gerry2005introductory}. On the other hand, it is well known that they do not pass the anticorrelation test and thus cannot directly represent a single-photon source \cite{aspect2013first}. In fact, observation of two-photon bunching and the HOM effect have a close connection with the anticorrelation test. One can observe these phenomena with sources that can pass the anticorrelation test. However, we have shown that it is possible to use quasi-classical states for observation of pure quantum mechanical effects. Our method which is based on randomization of the relative phase of these states in an interferometric setup can be exploited to probe quantum phenomena. This approach has been employed to reproduce the two-photon bunching and its complementary HOM effect as well as to test the Clauser, Horne, Shimony, and Holt (CHSH) form of the Bell inequality  \cite{arruda2020observation,mahdavifar2021violating}.

In a physical system, the existence of additional Degrees of Freedom (DoFs) enables a higher capability of information processing. In earlier works,  polarization has been the common DoF of light to investigate an optical system, but following the work of Allen et al. \cite{allen1992orbital}, the Orbital Angular Momentum (OAM) of light was added. Since this addition provides a higher dimensional Hilbert space, it enhances the capacity of information processing \cite{mair2001entanglement}. The quantum states in this d-dimensional Hilbert space are known as qudits versus qubits which only include two-dimensional subspaces. This allows a high-dimensional storage of quantum information on  a single photon. It has been shown that such generalization provides a vast range of theoretical extensions and experimental applications \cite{nielsen2002quantum,erhard2018twisted,yao2011orbital,padgett2017orbital,shen2019optical}. It empowers a higher information capacity, enhanced security, lower disturbance, stronger support for the entanglement against local realistic theories as well as better efficiency in quantum computing \cite{bechmann2000quantum,cerf2002security,mirhosseini2015high,sasaki2014practical,kaszlikowski2000violations,collins2002bell,dada2011experimental,campbell2012magic}. Examples employing OAM as a suitable DoF alone \cite{zhang2016engineering} or alongside other DoFs such as polarization are frequent. For instance, hyper-entangled states involve different physical variables such as polarization, OAM, etc, and are entangled in each of those DoFs \cite{barreiro2005generation,graham2015superdense}.

In this work, we present to our knowledge the first experimental observation of two-photon bunching with quasi-classical states of radiation carrying OAM. This is realized by controlling the total state through polarization and OAM from a continuous wave (CW) source. In our previous projects, the signal did not carry OAM and polarization was the only DoF involved \cite{arruda2020observation,mahdavifar2021violating}. One immediate benefit of this addition is the possibility of probing multi-photon interference based on the sorting approach of spatial modes of radiation. The idea is to sort different OAM states and then look at the sorted statistics of the interference \cite{rafsanjani2019sorting}. Non-linear sources face low photon numbers. For example, the rate of pair production for a few miliwatt of power from a pump source delivered to the non-linear crystal is about $10^{5}-10^{6}$ \cite{torres2011twisted}. This deficiency is a serious issue when further optical applications are needed, especially when one deals with the spatial properties of photons \cite{walborn2010spatial}. It can exceedingly challenge the outcome of the experiment. Here, we aim to show that quasi-classical states from a CW source can overcome this serious obstacle. For the same power from the CW source in our experiment, the photon rate is much more efficient. In the current work, we try to find the quantum signature of transferred information through these states carrying OAM. This method of generating OAM from a CW source paves the path to achieve multi-photon interference with higher efficiency. Also, the realization of hyperentanglement is another asset of this extension. In the following, we first explain the known theoretical background of the dynamics of quasi-classical states, then we look carefully at the details of our experiment and present our results.

From the dynamics of quasi-classical states, we can observe and find interesting facts. By definition, they are eigenstates of annihilation operator,
\begin{equation}
\hat{a}\ket{\alpha}=\alpha \ket{\alpha},
\end{equation}
in terms of number state, the coherent state $\ket{\alpha}$ is
\begin{equation}
\ket{\alpha}=\exp(-\frac{1}{2}|\alpha|^2)\sum_{n=0}^{\infty}\frac{\alpha^n}{\sqrt{n!}}\ket{n},
\label{cn}
\end{equation}
It is clear from Eq. (\ref{cn}) that quasi-classical states do not represent single-photon sources since these states are the superposition of different particle states. The probability distribution of these states follows Poisson distribution
\begin{equation}
P(m)=e^{-|\alpha|^2}\frac{|\alpha|^{2m}}{m!}=e^{-\mu}\frac{{\mu}^m}{m!},
\end{equation}
where
\begin{align}
\bra{\alpha}\hat{n}\ket{\alpha}&=\bra{\alpha}\hat{a}^{\dagger}\hat{a}\ket{\alpha}\notag\\
&=|\alpha|^{2}=\mu
\end{align}
is the average photon number over coherent states.
 
\begin{figure}
\centering
\includegraphics[width=.53\linewidth]{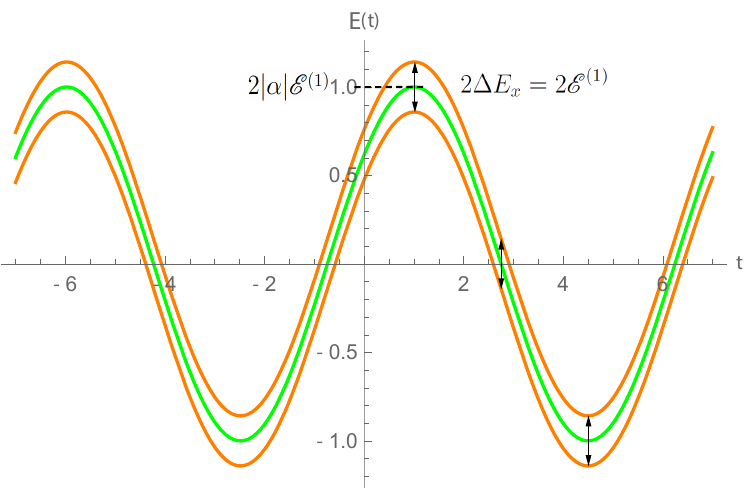}
\caption{Average electric field in the case of the coherent state with the uncertainty that is equal to the uncertainty of the vacuum state.}
\label{average}
\end{figure}

A specific line that we want to emphasize is the validity of particle-nature of radiation using these states, i.e., the limit of quantum observations predicted by such states. The quantity of interest is the average electric field over these states for a monochromatic light that propagates along the z direction with \textbf{E} and \textbf{B} fields oscillating along x and y, respectively. The electric field in the quantized frame with the frequency $\omega$ and the wave vector $\textbf{k}$ ($\omega/k=c$) is
\begin{equation}
\hat{{E}}_{x}(\textbf{r},t)=i\left( \frac{\hbar\omega}{2{\epsilon}_{0}{V}}\right)^{\frac{1}{2}} [\hat{a}~e^{i(\textbf{k}.\textbf{r}-{\omega}t)}-\hat{a}^{\dagger}e^{-i(\textbf{k}.\textbf{r}-{\omega}t)}],
\end{equation}
where $V$ is the volume of quantization. While clearly the average field is zero over the number state $\ket{n}$, the coherent state $\ket{\alpha}$ will produce what one observes in reality
\begin{align}
\bra{\alpha}\hat{E}_{x}(\textbf{r},t)\ket{\alpha}&=i\left( \frac{\hbar\omega}{2{\epsilon}_{0}{V}}\right)^\frac{1}{2}[{\alpha}~e^{i(\textbf{k}.\textbf{r}-{\omega}t)}-\alpha^{*}e^{-i(\textbf{k}.\textbf{r}-{\omega}t)}]\notag\\
&=2|\alpha|\left( \frac{\hbar\omega}{2{\epsilon}_{0}{V}}\right)^\frac{1}{2} sin({\omega}t-\textbf{k}.\textbf{r}-\theta)
\end{align}
in the last line we represent the coherent state in the polar form where $\ket{\alpha}=|\alpha|e^{i\theta}$. The uncertainty of the electric field over the coherent state
\begin{equation}
\Delta{\hat{E}_x}=\left( \frac{\hbar\omega}{2{\epsilon}_{0}{V}}\right)^\frac{1}{2}\equiv\mathscr{E}^{(1)}
\end{equation}
which is equal to the uncertainty calculated from the vacuum state $\ket{0}$, and $\mathscr{E}^{(1)}$ is single-photon amplitude of the electric field. The behavior of this mean field is plotted in Fig. \ref{average}. Now, the main point that we want to look upon is the domain of validity of the quantized picture. For this purpose, we consider the ratio of the uncertainty to the amplitude of the average field,
\begin{equation}
\frac{\Delta{\hat{E}_x}}{|\alpha|\mathscr{E}^{(1)}}=\frac{1}{|\alpha|}
=\frac{1}{\sqrt{\mu}}.
\label{ratio}
\end{equation}
 
For large values of $\mu$, as expected, this ratio tends to zero, indicating a classical limit. However, a peculiar behavior arises when the average photon number $\mu$ becomes less than one. For example, with $\mu = 10^{-2}$, the ratio is 10, and for $\mu = 10^{-4}$, this ratio increases to 100. In the limit of $\mu\ll1$, the average field is literally zero. Here, we emphasize that the significance of quantum effects becomes prominent and evident at low values of $\mu$. It is at this level that quantum mechanics emerges, even though the quantum nature of radiation is inherently present at all times. Previously, we mentioned that the average field over the number state, i.e., $\langle \hat{E} \rangle_{n}$ is zero, and now we observe from the dynamics of the coherent states that this fact manifests itself at the limit of $\mu\ll1$. This observation is somehow contradictory; either Quantum Fields Theory (QFT) does not produce the correct results, or its domain of application is limited. This means that considering light specifically as the stream of photons is not entirely correct before the interaction with matter or the attenuated limit. This condition suggests that the QFT is an effective theory, and the quantization frame is an approximation. Another facet of the quatized frame that one can think of to keep this notion globally is the possibility of other DoFs that exist within the structure of radiation, which we have not yet considered and observed. Another possibility is to state that the space spanned by the number state representation is bounded by the quantum realm, while the overall behavior of radiation can be described in a larger space, such as the space of coherent states and consequently keeping the quantization viewpoint in place. We think that this last idea is more relevant, considering the correct behavior of the mean field through the quantization picture over the quasi-classical states. Here, we do not aim to discuss the phenomenological aspects of physical reality. Rather, we simply wanted to mention some scenarios pertaining to the nature of light. Now returning to our main point, from Eq. (\ref{ratio}), one can contemplate the quantum nature of radiation. This implies that an observer can perceive the quantum aspect of radiation by approaching this weak limit without any interaction with matter. In this limit, the observer discovers that most of the radiation is devoid of photons, indicating that the dominant contribution arises from the vacuum state. For instance, with $\mu=10^{-2}$, the possibility of the single-photon state $\ket{1}$ is $10^{-2}$ relative to the vacuum $\ket{0}$, and possibility of observing two-photon state $\ket{2}$ is approximately $10^{-4}$ relative to the vacuum state. Now, depending on the precision that we want to reach, we can decouple our source from the optical system and achieve the desired $\mu$. The only limitation to achieve a higher photon rate in this approach is the coherence time and coherence length of the source. A shorter coherence time corresponds to a higher photon rate. It is worth noting that since there is no interaction with matter through nonlinear processes like SPDC, in this approach the efficiency will be higher compared to other sources that rely on non-linearity for generating single-photon states. This is the case because we reach the quantum limit only by decoupling of the source from the optical system. Now, there is an issue that needs to be addressed and a suitable solution needs to be found. Indeed, the quantum regime attainable through the weak limit is still not a perfect single-photon state. We have a mixed state with Poisson probability distribution. We need to isolate and separate the single-photon state. One method of separation that allows this is building an interferometric system. However, we will encounter a significant challenge in that case. The particle nature of radiation is unknown in the interferometric setting, and we cannot proceed to the measurement on the photon events unless we can resolve and address this problem. In fact, the solution is simple but profound, and it comes from the foundation of quantum mechanics itself. Introducing a temporal randomization in the interferometric setting and having a monochromatic radiation will pave the way, and it yields a consistent reading of the photon events. The details of this process will be discussed in the following lines. Next, we will describe our experimental setup, then develop the details of our approach and lastly present the results.
\begin{figure}
\centering
\includegraphics[scale=0.38]{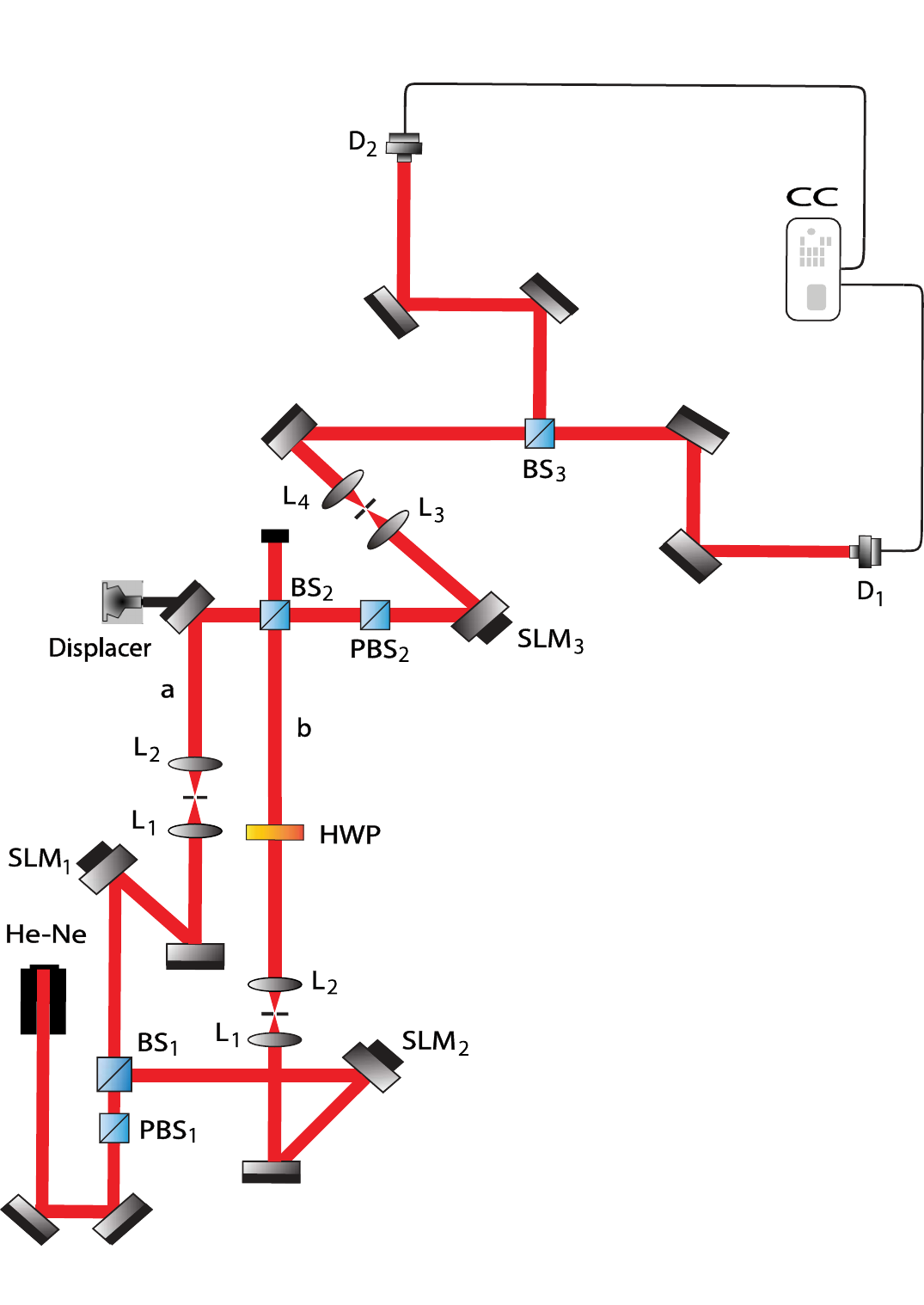}
\caption{Experimental setup for observation of two-photon bunching using quasi-classical states carrying OAM. Polarized Beam Splitter (PBS), Beam Splitter (BS), Spatial Light Modulater (SLM), Lens (L), Detector (D), Coincidence Counter (CC).}
\label{1}
\end{figure}

\begin{figure}
\centering
\includegraphics[scale=.151]{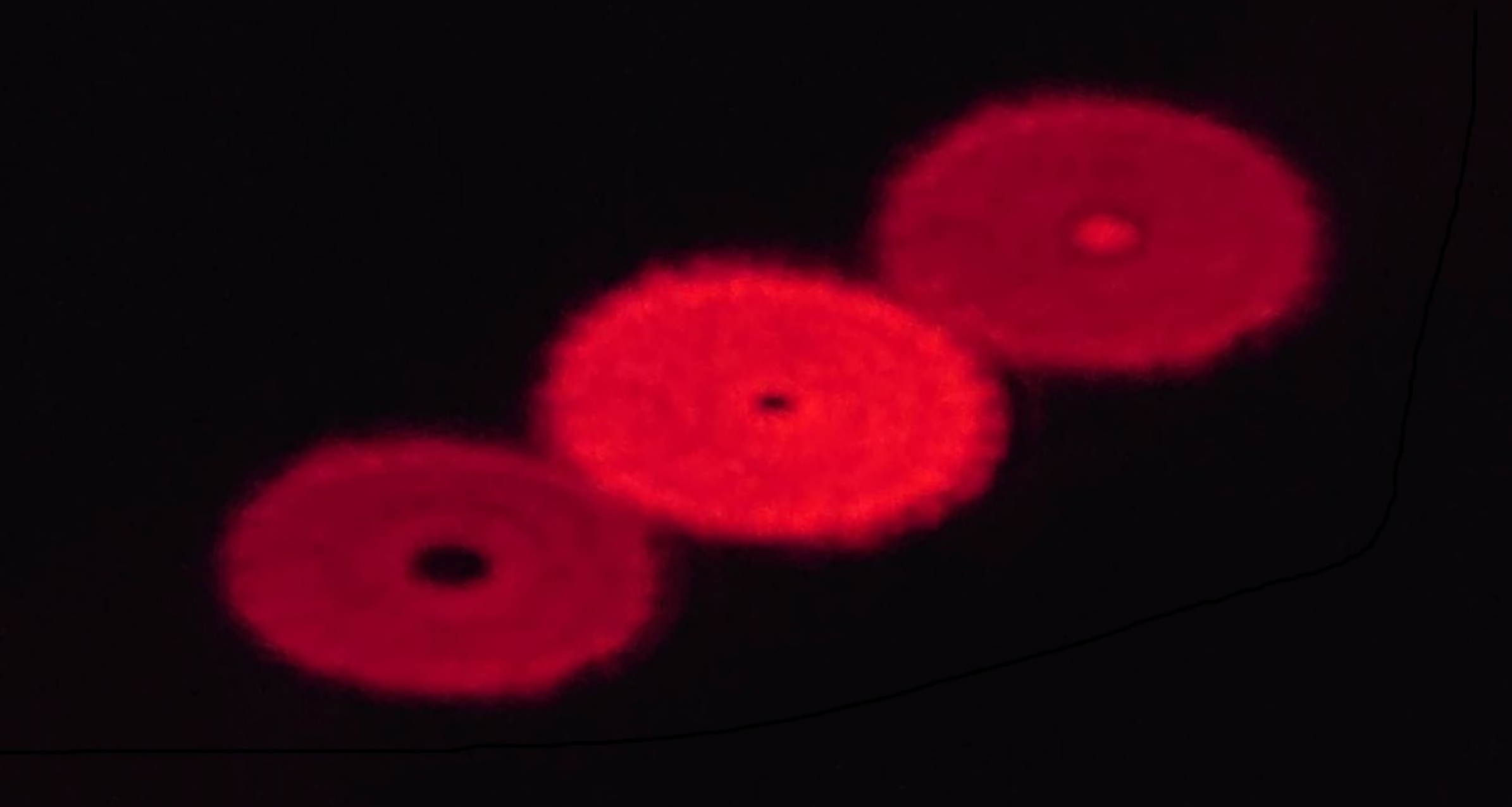}
\caption{Superposition of beams carrying OAM after SLM$_3$. The zero OAM state is observed as a result of the superposition of interfering beams carrying $l_{1}=l_{2}=1$ with $l_{3}=-1$.}
\label{superposition}
\end{figure}

The experimental setup is illustrated in Fig. \ref{1}. A Mach-Zehnder type interferometer is built involving two identical spatial light modulators (SLM$_{1}$ and SLM$_{2}$) as the projectors of photon states after reflecting the light from the source. The experiment implements a He-Ne laser coupled with the single-mode fiber as a source of quasi-classical modes. After collimation of the beam, to achieve a uniform polarization, we use a polarized beam splitter (PBS$_{1}$). Then the first balanced beam splitter (BS$_{1}$) is placed to split radiation into two beams for projection on the first and the second SLMs. These SLMs flatten the wavefront and project the same OAM quanta, for instance, `$l$'. Through identical lenses and arm lengths we bring the two beams together at the position of the second balanced beam splitter (BS$_{2}$). We thus have the interference pattern with two beams of the same polarization and OAM. Although in the figure the arm lengths may appear different, we have made sure that they have the same length (aspect ratio of the arms $a$ and $b$ is not preserved). After BS$_{2}$ we project one of the interfering beams to the third SLM (SLM$_{3}$) loaded with the opposite OAM, i.e., `$-l$'. We look for the result of this superposition which has zero OAM, to be observed at the first diffraction order as in Fig. \ref{superposition}. We can then transfer this signal after splitting it by the third balanced beam splitter (BS$_{3}$) to the single-photon avalanche detectors (D$_{1}$ and D$_{2}$) using either a pinhole/aperture and multi-mode fibers or the single-mode fibers that act like filters to only pass the basic Gaussian mode. The photon rate coming out of the single-mode fiber as a result of OAM superposition from a few miliwatt of power delivered to the interferometer is about $10^{11}-10^{12}$ which is much higher than the photon rate from the SPDC. It is also possible to enhance this rate slightly further with a more efficient alignment and optical manipulation. In fact, We have reached the interference with a visibility over 99$\%$ through the single-mode fiber after adding OAM. This fact supports the previous statement that with additional DoFs in a physical system, we gain a better ground for information processing.
One critical component in our setup is a displacer in arm $a$ that introduces a phase shift about $200\pi$ to the beam in this path. It creates a uniform temporal phase randomization between interfering beams which plays a central role in the experiment as we shall explain shortly. This delay in one arm is much smaller than the coherence length of the source to ensure the indistinguishability of the interfering photons. The resultant signals known as Phase Randomized Coherent States (PRCSs) are the main source of our investigation. We have a considerable photon rate, but the quantum effects will be observed in the weak limit. Hence, we decouple the source to reduce the intensity of radiation delivered to the interferometer. After this step, we call the radiation Phase Randomized Weak Coherent States (PRWCSs).

As mentioned above, coherent states cannot directly produce the quantum mechanical effects observed by single photon sources such as two-photon bunching or entanglement.  In the anticorrelation test, the radiation is sent to a beam splitter (BS); if the signal is detected in only one of the output ports, then the source is a single-photon source. This test does not allow quasi-classical states to represent single-photon sources because there is a nonzero probability of joint detection. Nevertheless, we deduced from our previous work the viability of observing purely quantum mechanical effects using these states. In this approach, we introduce a uniform time delay to one of the interfering beams. In an interferometric setting, the phase randomization process of the coherent state $\ket{\alpha}=\ket{\sqrt{\mu}e^{i\theta}}$ with the mean photon number $\mu=|\alpha|^{2}$ makes it possible to integrate over many temporal modes to ensure that the state of this signal is a mixed state with the Poisson distribution $P_{n}$ in the number states (Fock states) representation
\begin{equation}
\rho=\int{\frac{d{\theta}}{2\pi}}\ket{\sqrt{\mu}e^{i\theta}}\bra{\sqrt{\mu}e^{i\theta}}=\sum_{n=0}{P_n}\ket{n}\bra{n}.
\label{eq1}
\end{equation}

However, this is an emergent behavior that comes from the foundation of quantum mechanics itself, i.e., uncertainty principle. By randomizing the temporal phase, we gain a stable measure on the energy content of the propagating beams through the interferometer. Since our radiation is monochromatic, then the uncertainty of energy-time evolves to a number state-phase uncertainty,
\begin{equation}
\Delta{E}\Delta{t}={\hbar}\quad\Longrightarrow \quad \Delta{n}\Delta{\phi_t}=1\label{eq2}
\end{equation}
which means that upon the act of randomization in the temporal phase ${\phi_t}$, we achieve a reliable reading of the photon counts from the interferometer. In the absence of this action, it is not possible to gather consistent data. Once this randomization is employed, then we can filter the shares of higher-order photon numbers coming from each arm. This means that one measures the two-photon events while both arms $a$ and $b$ are open, and since the state is mixed, one also needs to consider the possibility of photon pairs originating from each specific arm. We deal with this situation by finding separate contributions, i.e., photon pairs only coming from arm $a$, by closing arm $b$, and then measuring two-photon counts only from arm $b$ while arm $a$ is closed. In this manner, we do three measurements for each data point and then subtract the second and the third counts from the first one (both arms being open) to retain the single-photon contributions to the interference. At this stage, it is necessary to point out some facts regarding the role of phase randomization. It is clear that we have the particle nature of radiation once there is no interference, i.e., radiation comes only from arm $a$ or $b$. In these two cases, we do not need any randomization, and the act of measurement will integrate over the possible modes and Eq. (\ref{eq1}) is satisfied. However, we do not have a meaningful measure on the number states in the case of both arms being open and having interference. Therefore, phase-randomization in this case is essential and we gain a stable measure of the number state due to this process. In this regard, the uncertainty principle as in Eq. (\ref{eq2}) is the underlying basis for the realization of Eq. (\ref{eq1}).

The initial state before BS$_{2}$ is
\begin{equation}
\ket{\Psi}=\ket{\psi_{S}}\otimes\ket{\psi_{L}},
\end{equation}
where $\ket{\psi_{S}}$ and $\ket{\psi_{L}}$ are the state functions of the polarization and OAM subspaces, respectively \cite{weihs2001photon},
\begin{equation}
\ket{\psi_{S}}=\frac{1}{\sqrt{2}}\left( \ket{1_{{a_1}H},1_{{b_2}H}}+e^{2i\phi}\ket{1_{{a_2}H},1_{{b_1}H}}\right) 
\end{equation}
and
\begin{equation}
\ket{\psi_{L}}=\frac{1}{\sqrt{2}}\left( \ket{1_{{a_1}l},1_{{b_2}l}}+e^{2il\phi}\ket{1_{{a_2}l},1_{{b_1}l}}\right),
\end{equation}
where $\ket{1_{{a_1}H},1_{{b_2}H}}$ denotes one-particle states of light with the horizontal polarization coming from path $a$ and $b$, respectively. Angle $\phi$ is the phase difference between arms $a$ and $b$. After the second BS, the states of interfering beams in each subspace become
\begin{align}\label{<4>}
\ket{\psi_{S}}&=ie^{i\phi}[(\ket{1_{{d_1}H},1_{{c_2}H}}-\ket{1_{{c_1}H},1_{{d_2}H}})\sin{\phi}\notag \\
&+(\ket{1_{{c_1}H},1_{{c_2}H}}+\ket{1_{{d_1}H},1_{{d_2}H}})\cos{\phi}],
\end{align}
and
\begin{align}\label{5}
\ket{\psi_{L}}&=ie^{il\phi}[(\ket{1_{{d_1}l},1_{{c_2}l}}-\ket{1_{{c_1}l},1_{{d_2}l}})\sin({l\phi})\notag \\
&+(\ket{1_{{c_1}l},1_{{c_2}l}}+\ket{1_{{d_1}l},1_{{d_2}l}})\cos({l\phi})].
\end{align}
We utilize Laguerre-Gaussian (LG) modes having OAM dependence of the form $\exp({il\phi})$ and each photon carrying $l\hbar$ quanta of OAM. The probability of two photons being detected in the same output ports c and d is given by the second terms in each subspace corresponding to Eqs. (\ref{<4>}) and (\ref{5}), respectively. Therefore, the probability of having two-photon bunching at each port c(d) is
\begin{equation}
P_{c_{1}c_{2}}=P_{d_{1}d_{2}}=\frac{1}{2}\cos^{2}\phi=\frac{1}{4}[1+\cos{(2\phi)}]~;~~l=1.
\end{equation}
In this work, the OAM quanta from SLM$_{1}$ and SLM$_{2}$ were fixed at $l_{1}=l_{2}=l=1$, and the state containing this OAM number was projected to the SLM$_{3}$ generating OAM `$-1$' i.e., $l_{3}=-1$. Thereby we can use the superposition of these states to get the resultant signal with zero OAM and within the first diffraction order after SLM$_{3}$. However, it is not necessary to have these values for the OAM. Indeed, as can be seen from state $\ket{\psi_{L}}$, if one knows what the correct order of diffraction is, then with any arbitrary OAM state, the probability of having two-photon coalescence comes from the second term with the scaled phase in $\cosine$ function as $\cos(l\phi)$.  Despite that, for $l=1$ the probability is the same in both subspaces. In the experiment, since we aim to filter the OAM states through superposition, then it is clear that the right choice is to have $l_{1}=l_{2}=-l_{3}$. As noted, having the value of OAM as $|l|=1$ will produce the same probability in both polarization and OAM subspaces.

The density operator representing the mixed state before BS$_{2}$ is
\begin{equation}
\rho = \sum_{i,j} P_{a}(i)P_{b}(j) \ket{i_{al_{1}},j_{bl_{2}}}\bra{i_{al_{1}},j_{bl_{2}}},
\end{equation}
where the probabilities $P_a(i), P_b(j)$ are Poisson distributions
\begin{equation}
P_s(n) = \frac{\mu_s^n}{n!}e^{-\mu_s}, \hspace{0.5 cm} s = a,b.
\end{equation}
Here, $\mu_a, \mu_b$ are the mean photon numbers at each input port of BS$_{2}$. State $\ket{i_{al_{1}},j_{bl_{2}}}$ describes an eigenstate with $i(j)$ photon number of OAM $l_{1}(l_{2})$ at the input port $a(b)$.
The goal is to separate two-photon events, among which we are merely interested in two-photon counts either at $c$ or $d$ ports.
The two-photon events at port $c$ are
\begin{align}
\mathcal{N}_{i,j}({\mu_a,\mu_b})&=\mu_a\mu_bP(0,2_c|1_a,1_b)+\frac{\mu_a^2}{2!} P(0,2_c|2_a,0_b)\notag \\
&+\frac{\mu_b^2}{2!}P(0,2_c| 0_a,2_b)+\mathcal{O}(\mu^3).
\end{align}\label{4}
The first term on the right-hand side is related to the desired state, and the other two terms are unwanted. The contributions to the latter are coming from
\begin{align}\nonumber
\mathcal{N}_{i,j}(\mu_a,0) =  \frac{\mu_a^2}{2}  P_{i,j}(0,1_c|2_a,0_b)+\mathcal{O}(\mu^3), \\ 
\mathcal{N}_{i,j}(0,\mu_b) =  \frac{\mu_b^2}{2}  P_{i,j}(0,2_c|0_a,2_b)+\mathcal{O}(\mu^3).\label{a0,0b}
\end{align}
As argued earlier, one needs to remove these two terms from the signal which leaves the pure contribution from the desired state
\begin{align}
\mu_a\mu_b P_{i,j}(0,2_c|1_a,1_b) & \simeq \mathcal{N}_{i,j}(\mu_a,\mu_b) - \mathcal{N}_{i,j}(\mu_a,0)\notag \\
&- \mathcal{N}_{i,j}(0,\mu_b).\label{ab,cd}
\end{align}
To justify this approximation, it is  important to mention that since the beam's intensity is low, the higher order terms, $\mathcal{O}(\mu^3)$, can be ignored. Let us address an obvious limitation of this method. The approximation in Eq. (\ref{ab,cd}) is valid for a weak source. It was noted that the photon rate out of a single-mode fiber after the superposition of OAM states is about $10^{11}-10^{12}$. On the other hand, the coherence time of the source is $1ns$. Therefore, with the $\mu=10^{-2}$, PRWCSs can be safely practiced resulting in a photon rate of $10^7$. However, suppose we have to build more complex optical systems; then definitely this approach shines in photon production and establishes its potential in high photon loss scenarios.

\begin{figure}
\centering
\includegraphics[scale=0.63]{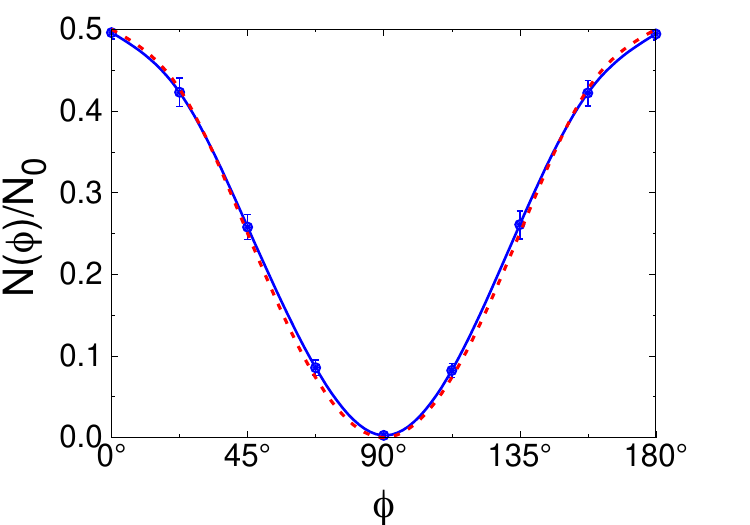}
\caption{Experimental realization of two-photon bunching which shows normalized counts versus polarization mismatched angle $\phi$ in $22.5^{\circ}$ steps.}
\label{2}
\end{figure}

\begin{figure}
\centering
\includegraphics[scale=0.61]{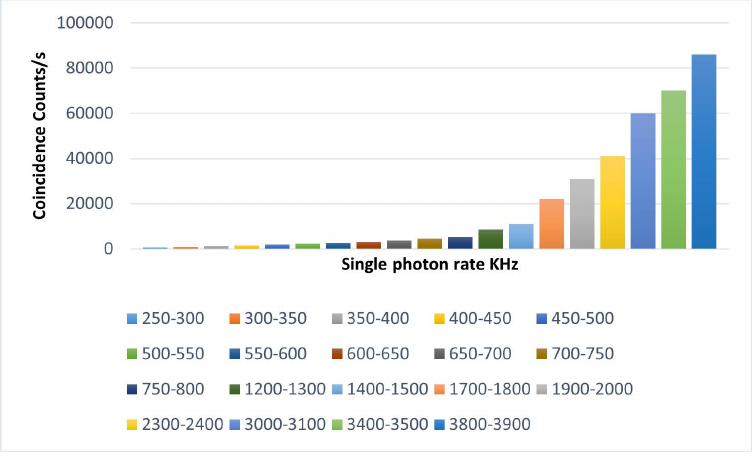}
\caption{Coincidence counts as a function of single-photon rate in unit of KHz.}
\label{cc}
\end{figure}

After explaining the concept and the math behind the PRWCSs, we are at the stage to present the results confirming the two-photon bunching using the coherent states carrying OAM. The data has been taken using polarization and OAM control of the signal. A half-wave plate (HWP) is inserted in arm $b$ of the interferometer to vary the polarization phase mismatch $\phi$ between two interfering beams, and OAM control occurs through superposition and the filtering process. Through these steps, we can check for the desired phenomenon. To clear the information about polarization of the interfering beams, we include another polarized beam splitter (PBS$_{2}$) before SLM$_{3}$. The data showing two-photon bunching is presented in Fig. \ref{2}. Here, the red dashed curve is theoretical prediction, and the solid blue curve is the best fit to the experimental data. Each error bar shows one standard deviation from the mean at each experimental data point. Furthermore, we aim to present information regarding the photon rates associated with the experiment. In both previous work \cite{arruda2020observation,mahdavifar2021violating}
the photon statistics were at a very low rate which means we explored up to $4\times10^{5}$ Hz single photon rate in the first project and up to $2\times10^{5}$ Hz for the second one. In the current experiment for the single rate of $4\times10^{5}$ Hz, we have approximately 2200 overall coincidence counts (joint detections) per second when both arms are open, prior to subtracting unwanted states. After the subtraction of the unwanted states, we have approximately 1300 coincidence counts per second as the absolute counts. The data acquisition period is 5s and over the coincidence window of 10ns. In the current work, we have pushed the single rate to the limit of our detector at $1.5\times10^{6}$ photons per second for a long usage, resulting in the double detection about 26000/s, and the absolute counts regarding the pure contributions of the single-photons around 11000/s. Also, it is possible to use our detector for a short time and go up to 4MHz. At this limit, we have more than 85000/s absolute counts. This happens at the maxima of our result corresponding to $0^{\circ}$ or $180^{\circ}$. However, with the aforementioned source, we can enhance the single rate up to 30MHz with just 2 percent error in the Eq. (\ref{ab,cd}). To the best of our knowledge, this result is the highest coincidence rate observed in similar settings on two-photon bunching and the HOM effect. In the case of having a source with shorter coherence time, it is possible to further increase this rate, potentially surpassing the current photon rate of commercially available single-photon detectors. Therefore, we think that this approach has the potential to advance our current capabilities and opens up new opportunities for exploring quantum phenomena. In Fig. \ref{cc}, we have plotted the coincidence counts versus the average single-photon rate in the unit of KHz. Our intuition is that this increment in the absolute coincidence counts will be saturated as we increase the single-photon rate, and this occurs when we have the contribution from higher order terms. 
At that limit, a CW source no longer can be used to extract a single-photon contribution. This is the only fundamental limitation of this approach and as stated, by having a source with the shorter coherence time one can enhance the single and double photodetections rate.

In conclusion, what we have observed with quasi-classical states carrying OAM is the experimental realization of the quantum behavior of such states, reproducing one of the basic phenomena in the domain of quantum optics, namely two-photon bunching with a considerable coincidence rate and the possibility to enhance it further. We have also experimentally observed the existence of the number state-phase uncertainty, even though there is currently no operator representing the phase. The standard extension of this toolbox is that the PRWCSs can serve to investigate manifold quantum aspects of light pertinent to modern quantum technologies as well as to search for a deeper understanding of fundamental physics. To give one  example, it is possible to probe multi-photon interference. Since quasi-classical states do not have any shortage in photon numbers, they do not encounter the scalability problem as much as the other single-photon sources do, implying that this technique is much more robust against photon loss. This can be done by an efficient sorting approach to the OAM states and then looking for the multi-photon interference. Another example is the use of the PRWCSs to realize entangled based protocols of the quantum key distribution (QKD) \cite{ekert1991quantum}. Since the violation of the Bell inequality with such states has successfully been verified, the existence of an eavesdropper in the path of communication will be determined if the Bell inequality holds. The current work with OAM provides even higher security to the QKD protocols. Moreover, the generation of hyperentangled states using this approach is quite straightforward, which makes it possible to exploit these states in quantum communication protocols. A CW source is more cost-effective than a pulsed source, and this is another advantage of using such a method in the domain of quantum optics and quantum information sciences. Now that we have one more experimental confirmation involving quasi-classical states carrying OAM, a more diverse set of experimental setups can be built to test the validity and the limitations of such states in exploring the quantum world.
\section*{Acknowledgements}
The author thank Olga Korotkova, S. M. Hashemi Rafsanjani and Miles Padgett for their helpful comments.

%
%
\section*{Conflict of Interests}
The author declare no conflict of interests.


\begin{thebibliography}{37}%
\makeatletter
\providecommand \@ifxundefined [1]{%
 \@ifx{#1\undefined}
}%
\providecommand \@ifnum [1]{%
 \ifnum #1\expandafter \@firstoftwo
 \else \expandafter \@secondoftwo
 \fi
}%
\providecommand \@ifx [1]{%
 \ifx #1\expandafter \@firstoftwo
 \else \expandafter \@secondoftwo
 \fi
}%
\providecommand \natexlab [1]{#1}%
\providecommand \enquote  [1]{``#1''}%
\providecommand \bibnamefont  [1]{#1}%
\providecommand \bibfnamefont [1]{#1}%
\providecommand \citenamefont [1]{#1}%
\providecommand \href@noop [0]{\@secondoftwo}%
\providecommand \href [0]{\begingroup \@sanitize@url \@href}%
\providecommand \@href[1]{\@@startlink{#1}\@@href}%
\providecommand \@@href[1]{\endgroup#1\@@endlink}%
\providecommand \@sanitize@url [0]{\catcode `\\12\catcode `\$12\catcode
  `\&12\catcode `\#12\catcode `\^12\catcode `\_12\catcode `\%12\relax}%
\providecommand \@@startlink[1]{}%
\providecommand \@@endlink[0]{}%
\providecommand \url  [0]{\begingroup\@sanitize@url \@url }%
\providecommand \@url [1]{\endgroup\@href {#1}{\urlprefix }}%
\providecommand \urlprefix  [0]{URL }%
\providecommand \Eprint [0]{\href }%
\providecommand \doibase [0]{https://doi.org/}%
\providecommand \selectlanguage [0]{\@gobble}%
\providecommand \bibinfo  [0]{\@secondoftwo}%
\providecommand \bibfield  [0]{\@secondoftwo}%
\providecommand \translation [1]{[#1]}%
\providecommand \BibitemOpen [0]{}%
\providecommand \bibitemStop [0]{}%
\providecommand \bibitemNoStop [0]{.\EOS\space}%
\providecommand \EOS [0]{\spacefactor3000\relax}%
\providecommand \BibitemShut  [1]{\csname bibitem#1\endcsname}%
\let\auto@bib@innerbib\@empty
\bibitem [{\citenamefont {Hong}\ \emph {et~al.}(1987)\citenamefont {Hong},
  \citenamefont {Ou},\ and\ \citenamefont {Mandel}}]{hong1987measurement}%
  \BibitemOpen
  \bibfield  {author} {\bibinfo {author} {\bibfnamefont {C.-K.}\ \bibnamefont
  {Hong}}, \bibinfo {author} {\bibfnamefont {Z.-Y.}\ \bibnamefont {Ou}},\ and\
  \bibinfo {author} {\bibfnamefont {L.}~\bibnamefont {Mandel}},\ }\bibfield
  {title} {\bibinfo {title} {Measurement of subpicosecond time intervals
  between two photons by interference},\ }\href@noop {} {\bibfield  {journal}
  {\bibinfo  {journal} {Physical review letters}\ }\textbf {\bibinfo {volume}
  {59}},\ \bibinfo {pages} {2044} (\bibinfo {year} {1987})}\BibitemShut
  {NoStop}%
\bibitem [{\citenamefont {Bouwmeester}\ \emph {et~al.}(1997)\citenamefont
  {Bouwmeester}, \citenamefont {Pan}, \citenamefont {Mattle}, \citenamefont
  {Eibl}, \citenamefont {Weinfurter},\ and\ \citenamefont
  {Zeilinger}}]{bouwmeester1997experimental}%
  \BibitemOpen
  \bibfield  {author} {\bibinfo {author} {\bibfnamefont {D.}~\bibnamefont
  {Bouwmeester}}, \bibinfo {author} {\bibfnamefont {J.-W.}\ \bibnamefont
  {Pan}}, \bibinfo {author} {\bibfnamefont {K.}~\bibnamefont {Mattle}},
  \bibinfo {author} {\bibfnamefont {M.}~\bibnamefont {Eibl}}, \bibinfo {author}
  {\bibfnamefont {H.}~\bibnamefont {Weinfurter}},\ and\ \bibinfo {author}
  {\bibfnamefont {A.}~\bibnamefont {Zeilinger}},\ }\bibfield  {title} {\bibinfo
  {title} {Experimental quantum teleportation},\ }\href@noop {} {\bibfield
  {journal} {\bibinfo  {journal} {Nature}\ }\textbf {\bibinfo {volume} {390}},\
  \bibinfo {pages} {575} (\bibinfo {year} {1997})}\BibitemShut {NoStop}%
\bibitem [{\citenamefont {Knill}\ \emph {et~al.}(2001)\citenamefont {Knill},
  \citenamefont {Laflamme},\ and\ \citenamefont {Milburn}}]{knill2001scheme}%
  \BibitemOpen
  \bibfield  {author} {\bibinfo {author} {\bibfnamefont {E.}~\bibnamefont
  {Knill}}, \bibinfo {author} {\bibfnamefont {R.}~\bibnamefont {Laflamme}},\
  and\ \bibinfo {author} {\bibfnamefont {G.~J.}\ \bibnamefont {Milburn}},\
  }\bibfield  {title} {\bibinfo {title} {A scheme for efficient quantum
  computation with linear optics},\ }\href@noop {} {\bibfield  {journal}
  {\bibinfo  {journal} {nature}\ }\textbf {\bibinfo {volume} {409}},\ \bibinfo
  {pages} {46} (\bibinfo {year} {2001})}\BibitemShut {NoStop}%
\bibitem [{\citenamefont {Giovannetti}\ \emph {et~al.}(2011)\citenamefont
  {Giovannetti}, \citenamefont {Lloyd},\ and\ \citenamefont
  {Maccone}}]{giovannetti2011advances}%
  \BibitemOpen
  \bibfield  {author} {\bibinfo {author} {\bibfnamefont {V.}~\bibnamefont
  {Giovannetti}}, \bibinfo {author} {\bibfnamefont {S.}~\bibnamefont {Lloyd}},\
  and\ \bibinfo {author} {\bibfnamefont {L.}~\bibnamefont {Maccone}},\
  }\bibfield  {title} {\bibinfo {title} {Advances in quantum metrology},\
  }\href@noop {} {\bibfield  {journal} {\bibinfo  {journal} {Nature photonics}\
  }\textbf {\bibinfo {volume} {5}},\ \bibinfo {pages} {222} (\bibinfo {year}
  {2011})}\BibitemShut {NoStop}%
\bibitem [{\citenamefont {Boyd}(2020)}]{boyd2020nonlinear}%
  \BibitemOpen
  \bibfield  {author} {\bibinfo {author} {\bibfnamefont {R.~W.}\ \bibnamefont
  {Boyd}},\ }\href@noop {} {\emph {\bibinfo {title} {Nonlinear optics}}}\
  (\bibinfo  {publisher} {Academic press},\ \bibinfo {year} {2020})\BibitemShut
  {NoStop}%
\bibitem [{\citenamefont {Bachor}\ and\ \citenamefont
  {Ralph}(2019)}]{bachor2019guide}%
  \BibitemOpen
  \bibfield  {author} {\bibinfo {author} {\bibfnamefont {H.-A.}\ \bibnamefont
  {Bachor}}\ and\ \bibinfo {author} {\bibfnamefont {T.~C.}\ \bibnamefont
  {Ralph}},\ }\href@noop {} {\emph {\bibinfo {title} {A guide to experiments in
  quantum optics}}}\ (\bibinfo  {publisher} {John Wiley \& Sons},\ \bibinfo
  {year} {2019})\BibitemShut {NoStop}%
\bibitem [{\citenamefont {Eisaman}\ \emph {et~al.}(2011)\citenamefont
  {Eisaman}, \citenamefont {Fan}, \citenamefont {Migdall},\ and\ \citenamefont
  {Polyakov}}]{eisaman2011invited}%
  \BibitemOpen
  \bibfield  {author} {\bibinfo {author} {\bibfnamefont {M.~D.}\ \bibnamefont
  {Eisaman}}, \bibinfo {author} {\bibfnamefont {J.}~\bibnamefont {Fan}},
  \bibinfo {author} {\bibfnamefont {A.}~\bibnamefont {Migdall}},\ and\ \bibinfo
  {author} {\bibfnamefont {S.~V.}\ \bibnamefont {Polyakov}},\ }\bibfield
  {title} {\bibinfo {title} {Invited review article: Single-photon sources and
  detectors},\ }\href@noop {} {\bibfield  {journal} {\bibinfo  {journal}
  {Review of scientific instruments}\ }\textbf {\bibinfo {volume} {82}},\
  \bibinfo {pages} {071101} (\bibinfo {year} {2011})}\BibitemShut {NoStop}%
\bibitem [{\citenamefont {Sudarshan}(1963)}]{sudarshan1963equivalence}%
  \BibitemOpen
  \bibfield  {author} {\bibinfo {author} {\bibfnamefont {E.}~\bibnamefont
  {Sudarshan}},\ }\bibfield  {title} {\bibinfo {title} {Equivalence of
  semiclassical and quantum mechanical descriptions of statistical light
  beams},\ }\href@noop {} {\bibfield  {journal} {\bibinfo  {journal} {Physical
  Review Letters}\ }\textbf {\bibinfo {volume} {10}},\ \bibinfo {pages} {277}
  (\bibinfo {year} {1963})}\BibitemShut {NoStop}%
\bibitem [{\citenamefont {Glauber}(1963)}]{glauber1963coherent}%
  \BibitemOpen
  \bibfield  {author} {\bibinfo {author} {\bibfnamefont {R.~J.}\ \bibnamefont
  {Glauber}},\ }\bibfield  {title} {\bibinfo {title} {Coherent and incoherent
  states of the radiation field},\ }\href@noop {} {\bibfield  {journal}
  {\bibinfo  {journal} {Physical Review}\ }\textbf {\bibinfo {volume} {131}},\
  \bibinfo {pages} {2766} (\bibinfo {year} {1963})}\BibitemShut {NoStop}%
\bibitem [{\citenamefont {Grynberg}\ \emph {et~al.}(2010)\citenamefont
  {Grynberg}, \citenamefont {Aspect},\ and\ \citenamefont
  {Fabre}}]{grynberg2010introduction}%
  \BibitemOpen
  \bibfield  {author} {\bibinfo {author} {\bibfnamefont {G.}~\bibnamefont
  {Grynberg}}, \bibinfo {author} {\bibfnamefont {A.}~\bibnamefont {Aspect}},\
  and\ \bibinfo {author} {\bibfnamefont {C.}~\bibnamefont {Fabre}},\
  }\href@noop {} {\emph {\bibinfo {title} {Introduction to quantum optics: from
  the semi-classical approach to quantized light}}}\ (\bibinfo  {publisher}
  {Cambridge university press},\ \bibinfo {year} {2010})\BibitemShut {NoStop}%
\bibitem [{\citenamefont {Gerry}\ \emph {et~al.}(2005)\citenamefont {Gerry},
  \citenamefont {Knight},\ and\ \citenamefont
  {Knight}}]{gerry2005introductory}%
  \BibitemOpen
  \bibfield  {author} {\bibinfo {author} {\bibfnamefont {C.}~\bibnamefont
  {Gerry}}, \bibinfo {author} {\bibfnamefont {P.}~\bibnamefont {Knight}},\ and\
  \bibinfo {author} {\bibfnamefont {P.~L.}\ \bibnamefont {Knight}},\
  }\href@noop {} {\emph {\bibinfo {title} {Introductory quantum optics}}}\
  (\bibinfo  {publisher} {Cambridge university press},\ \bibinfo {year}
  {2005})\BibitemShut {NoStop}%
\bibitem [{\citenamefont {Aspect}\ and\ \citenamefont
  {Grangier}(2013)}]{aspect2013first}%
  \BibitemOpen
  \bibfield  {author} {\bibinfo {author} {\bibfnamefont {A.}~\bibnamefont
  {Aspect}}\ and\ \bibinfo {author} {\bibfnamefont {P.}~\bibnamefont
  {Grangier}},\ }\bibfield  {title} {\bibinfo {title} {The first single-photon
  sources},\ }\href@noop {} {\bibfield  {journal} {\bibinfo  {journal}
  {Single-Photon Generation and Detection}\ ,\ \bibinfo {pages} {315}}
  (\bibinfo {year} {2013})}\BibitemShut {NoStop}%
\bibitem [{\citenamefont {Arruda}\ \emph {et~al.}(2020)\citenamefont {Arruda},
  \citenamefont {Mahdavifar}, \citenamefont {Krop}, \citenamefont {Ribeiro},\
  and\ \citenamefont {Rafsanjani}}]{arruda2020observation}%
  \BibitemOpen
  \bibfield  {author} {\bibinfo {author} {\bibfnamefont {M.~F.}\ \bibnamefont
  {Arruda}}, \bibinfo {author} {\bibfnamefont {M.}~\bibnamefont {Mahdavifar}},
  \bibinfo {author} {\bibfnamefont {T.}~\bibnamefont {Krop}}, \bibinfo {author}
  {\bibfnamefont {P.~H.~S.}\ \bibnamefont {Ribeiro}},\ and\ \bibinfo {author}
  {\bibfnamefont {S.~M.~H.}\ \bibnamefont {Rafsanjani}},\ }\bibfield  {title}
  {\bibinfo {title} {Observation of two-photon coalescence in weak coherent
  wave packets},\ }\href@noop {} {\bibfield  {journal} {\bibinfo  {journal}
  {JOSA B}\ }\textbf {\bibinfo {volume} {37}},\ \bibinfo {pages} {2901}
  (\bibinfo {year} {2020})}\BibitemShut {NoStop}%
\bibitem [{\citenamefont {Mahdavifar}\ and\ \citenamefont
  {SM}(2021)}]{mahdavifar2021violating}%
  \BibitemOpen
  \bibfield  {author} {\bibinfo {author} {\bibfnamefont {M.}~\bibnamefont
  {Mahdavifar}}\ and\ \bibinfo {author} {\bibfnamefont {H.~R.}\ \bibnamefont
  {SM}},\ }\bibfield  {title} {\bibinfo {title} {Violating bell inequality
  using weak coherent states.},\ }\href@noop {} {\bibfield  {journal} {\bibinfo
   {journal} {Optics Letters}\ }\textbf {\bibinfo {volume} {46}},\ \bibinfo
  {pages} {5998} (\bibinfo {year} {2021})}\BibitemShut {NoStop}%
\bibitem [{\citenamefont {Allen}\ \emph {et~al.}(1992)\citenamefont {Allen},
  \citenamefont {Beijersbergen}, \citenamefont {Spreeuw},\ and\ \citenamefont
  {Woerdman}}]{allen1992orbital}%
  \BibitemOpen
  \bibfield  {author} {\bibinfo {author} {\bibfnamefont {L.}~\bibnamefont
  {Allen}}, \bibinfo {author} {\bibfnamefont {M.~W.}\ \bibnamefont
  {Beijersbergen}}, \bibinfo {author} {\bibfnamefont {R.}~\bibnamefont
  {Spreeuw}},\ and\ \bibinfo {author} {\bibfnamefont {J.}~\bibnamefont
  {Woerdman}},\ }\bibfield  {title} {\bibinfo {title} {Orbital angular momentum
  of light and the transformation of laguerre-gaussian laser modes},\
  }\href@noop {} {\bibfield  {journal} {\bibinfo  {journal} {Physical review
  A}\ }\textbf {\bibinfo {volume} {45}},\ \bibinfo {pages} {8185} (\bibinfo
  {year} {1992})}\BibitemShut {NoStop}%
\bibitem [{\citenamefont {Mair}\ \emph {et~al.}(2001)\citenamefont {Mair},
  \citenamefont {Vaziri}, \citenamefont {Weihs},\ and\ \citenamefont
  {Zeilinger}}]{mair2001entanglement}%
  \BibitemOpen
  \bibfield  {author} {\bibinfo {author} {\bibfnamefont {A.}~\bibnamefont
  {Mair}}, \bibinfo {author} {\bibfnamefont {A.}~\bibnamefont {Vaziri}},
  \bibinfo {author} {\bibfnamefont {G.}~\bibnamefont {Weihs}},\ and\ \bibinfo
  {author} {\bibfnamefont {A.}~\bibnamefont {Zeilinger}},\ }\bibfield  {title}
  {\bibinfo {title} {Entanglement of the orbital angular momentum states of
  photons},\ }\href@noop {} {\bibfield  {journal} {\bibinfo  {journal}
  {Nature}\ }\textbf {\bibinfo {volume} {412}},\ \bibinfo {pages} {313}
  (\bibinfo {year} {2001})}\BibitemShut {NoStop}%
\bibitem [{\citenamefont {Nielsen}\ and\ \citenamefont
  {Chuang}(2002)}]{nielsen2002quantum}%
  \BibitemOpen
  \bibfield  {author} {\bibinfo {author} {\bibfnamefont {M.~A.}\ \bibnamefont
  {Nielsen}}\ and\ \bibinfo {author} {\bibfnamefont {I.}~\bibnamefont
  {Chuang}},\ }\href@noop {} {\bibinfo {title} {Quantum computation and quantum
  information}} (\bibinfo {year} {2002})\BibitemShut {NoStop}%
\bibitem [{\citenamefont {Erhard}\ \emph {et~al.}(2018)\citenamefont {Erhard},
  \citenamefont {Fickler}, \citenamefont {Krenn},\ and\ \citenamefont
  {Zeilinger}}]{erhard2018twisted}%
  \BibitemOpen
  \bibfield  {author} {\bibinfo {author} {\bibfnamefont {M.}~\bibnamefont
  {Erhard}}, \bibinfo {author} {\bibfnamefont {R.}~\bibnamefont {Fickler}},
  \bibinfo {author} {\bibfnamefont {M.}~\bibnamefont {Krenn}},\ and\ \bibinfo
  {author} {\bibfnamefont {A.}~\bibnamefont {Zeilinger}},\ }\bibfield  {title}
  {\bibinfo {title} {Twisted photons: new quantum perspectives in high
  dimensions},\ }\href@noop {} {\bibfield  {journal} {\bibinfo  {journal}
  {Light: Science \& Applications}\ }\textbf {\bibinfo {volume} {7}},\ \bibinfo
  {pages} {17146} (\bibinfo {year} {2018})}\BibitemShut {NoStop}%
\bibitem [{\citenamefont {Yao}\ and\ \citenamefont
  {Padgett}(2011)}]{yao2011orbital}%
  \BibitemOpen
  \bibfield  {author} {\bibinfo {author} {\bibfnamefont {A.~M.}\ \bibnamefont
  {Yao}}\ and\ \bibinfo {author} {\bibfnamefont {M.~J.}\ \bibnamefont
  {Padgett}},\ }\bibfield  {title} {\bibinfo {title} {Orbital angular momentum:
  origins, behavior and applications},\ }\href@noop {} {\bibfield  {journal}
  {\bibinfo  {journal} {Advances in optics and photonics}\ }\textbf {\bibinfo
  {volume} {3}},\ \bibinfo {pages} {161} (\bibinfo {year} {2011})}\BibitemShut
  {NoStop}%
\bibitem [{\citenamefont {Padgett}(2017)}]{padgett2017orbital}%
  \BibitemOpen
  \bibfield  {author} {\bibinfo {author} {\bibfnamefont {M.~J.}\ \bibnamefont
  {Padgett}},\ }\bibfield  {title} {\bibinfo {title} {Orbital angular momentum
  25 years on},\ }\href@noop {} {\bibfield  {journal} {\bibinfo  {journal}
  {Optics express}\ }\textbf {\bibinfo {volume} {25}},\ \bibinfo {pages}
  {11265} (\bibinfo {year} {2017})}\BibitemShut {NoStop}%
\bibitem [{\citenamefont {Shen}\ \emph {et~al.}(2019)\citenamefont {Shen},
  \citenamefont {Wang}, \citenamefont {Xie}, \citenamefont {Min}, \citenamefont
  {Fu}, \citenamefont {Liu}, \citenamefont {Gong},\ and\ \citenamefont
  {Yuan}}]{shen2019optical}%
  \BibitemOpen
  \bibfield  {author} {\bibinfo {author} {\bibfnamefont {Y.}~\bibnamefont
  {Shen}}, \bibinfo {author} {\bibfnamefont {X.}~\bibnamefont {Wang}}, \bibinfo
  {author} {\bibfnamefont {Z.}~\bibnamefont {Xie}}, \bibinfo {author}
  {\bibfnamefont {C.}~\bibnamefont {Min}}, \bibinfo {author} {\bibfnamefont
  {X.}~\bibnamefont {Fu}}, \bibinfo {author} {\bibfnamefont {Q.}~\bibnamefont
  {Liu}}, \bibinfo {author} {\bibfnamefont {M.}~\bibnamefont {Gong}},\ and\
  \bibinfo {author} {\bibfnamefont {X.}~\bibnamefont {Yuan}},\ }\bibfield
  {title} {\bibinfo {title} {Optical vortices 30 years on: Oam manipulation
  from topological charge to multiple singularities},\ }\href@noop {}
  {\bibfield  {journal} {\bibinfo  {journal} {Light: Science \& Applications}\
  }\textbf {\bibinfo {volume} {8}},\ \bibinfo {pages} {1} (\bibinfo {year}
  {2019})}\BibitemShut {NoStop}%
\bibitem [{\citenamefont {Bechmann-Pasquinucci}\ and\ \citenamefont
  {Tittel}(2000)}]{bechmann2000quantum}%
  \BibitemOpen
  \bibfield  {author} {\bibinfo {author} {\bibfnamefont {H.}~\bibnamefont
  {Bechmann-Pasquinucci}}\ and\ \bibinfo {author} {\bibfnamefont
  {W.}~\bibnamefont {Tittel}},\ }\bibfield  {title} {\bibinfo {title} {Quantum
  cryptography using larger alphabets},\ }\href@noop {} {\bibfield  {journal}
  {\bibinfo  {journal} {Physical Review A}\ }\textbf {\bibinfo {volume} {61}},\
  \bibinfo {pages} {062308} (\bibinfo {year} {2000})}\BibitemShut {NoStop}%
\bibitem [{\citenamefont {Cerf}\ \emph {et~al.}(2002)\citenamefont {Cerf},
  \citenamefont {Bourennane}, \citenamefont {Karlsson},\ and\ \citenamefont
  {Gisin}}]{cerf2002security}%
  \BibitemOpen
  \bibfield  {author} {\bibinfo {author} {\bibfnamefont {N.~J.}\ \bibnamefont
  {Cerf}}, \bibinfo {author} {\bibfnamefont {M.}~\bibnamefont {Bourennane}},
  \bibinfo {author} {\bibfnamefont {A.}~\bibnamefont {Karlsson}},\ and\
  \bibinfo {author} {\bibfnamefont {N.}~\bibnamefont {Gisin}},\ }\bibfield
  {title} {\bibinfo {title} {Security of quantum key distribution using d-level
  systems},\ }\href@noop {} {\bibfield  {journal} {\bibinfo  {journal}
  {Physical review letters}\ }\textbf {\bibinfo {volume} {88}},\ \bibinfo
  {pages} {127902} (\bibinfo {year} {2002})}\BibitemShut {NoStop}%
\bibitem [{\citenamefont {Mirhosseini}\ \emph {et~al.}(2015)\citenamefont
  {Mirhosseini}, \citenamefont {Maga{\~n}a-Loaiza}, \citenamefont
  {O’Sullivan}, \citenamefont {Rodenburg}, \citenamefont {Malik},
  \citenamefont {Lavery}, \citenamefont {Padgett}, \citenamefont {Gauthier},\
  and\ \citenamefont {Boyd}}]{mirhosseini2015high}%
  \BibitemOpen
  \bibfield  {author} {\bibinfo {author} {\bibfnamefont {M.}~\bibnamefont
  {Mirhosseini}}, \bibinfo {author} {\bibfnamefont {O.~S.}\ \bibnamefont
  {Maga{\~n}a-Loaiza}}, \bibinfo {author} {\bibfnamefont {M.~N.}\ \bibnamefont
  {O’Sullivan}}, \bibinfo {author} {\bibfnamefont {B.}~\bibnamefont
  {Rodenburg}}, \bibinfo {author} {\bibfnamefont {M.}~\bibnamefont {Malik}},
  \bibinfo {author} {\bibfnamefont {M.~P.}\ \bibnamefont {Lavery}}, \bibinfo
  {author} {\bibfnamefont {M.~J.}\ \bibnamefont {Padgett}}, \bibinfo {author}
  {\bibfnamefont {D.~J.}\ \bibnamefont {Gauthier}},\ and\ \bibinfo {author}
  {\bibfnamefont {R.~W.}\ \bibnamefont {Boyd}},\ }\bibfield  {title} {\bibinfo
  {title} {High-dimensional quantum cryptography with twisted light},\
  }\href@noop {} {\bibfield  {journal} {\bibinfo  {journal} {New Journal of
  Physics}\ }\textbf {\bibinfo {volume} {17}},\ \bibinfo {pages} {033033}
  (\bibinfo {year} {2015})}\BibitemShut {NoStop}%
\bibitem [{\citenamefont {Sasaki}\ \emph {et~al.}(2014)\citenamefont {Sasaki},
  \citenamefont {Yamamoto},\ and\ \citenamefont
  {Koashi}}]{sasaki2014practical}%
  \BibitemOpen
  \bibfield  {author} {\bibinfo {author} {\bibfnamefont {T.}~\bibnamefont
  {Sasaki}}, \bibinfo {author} {\bibfnamefont {Y.}~\bibnamefont {Yamamoto}},\
  and\ \bibinfo {author} {\bibfnamefont {M.}~\bibnamefont {Koashi}},\
  }\bibfield  {title} {\bibinfo {title} {Practical quantum key distribution
  protocol without monitoring signal disturbance},\ }\href@noop {} {\bibfield
  {journal} {\bibinfo  {journal} {Nature}\ }\textbf {\bibinfo {volume} {509}},\
  \bibinfo {pages} {475} (\bibinfo {year} {2014})}\BibitemShut {NoStop}%
\bibitem [{\citenamefont {Kaszlikowski}\ \emph {et~al.}(2000)\citenamefont
  {Kaszlikowski}, \citenamefont {Gnaci{\'n}ski}, \citenamefont {{\.Z}ukowski},
  \citenamefont {Miklaszewski},\ and\ \citenamefont
  {Zeilinger}}]{kaszlikowski2000violations}%
  \BibitemOpen
  \bibfield  {author} {\bibinfo {author} {\bibfnamefont {D.}~\bibnamefont
  {Kaszlikowski}}, \bibinfo {author} {\bibfnamefont {P.}~\bibnamefont
  {Gnaci{\'n}ski}}, \bibinfo {author} {\bibfnamefont {M.}~\bibnamefont
  {{\.Z}ukowski}}, \bibinfo {author} {\bibfnamefont {W.}~\bibnamefont
  {Miklaszewski}},\ and\ \bibinfo {author} {\bibfnamefont {A.}~\bibnamefont
  {Zeilinger}},\ }\bibfield  {title} {\bibinfo {title} {Violations of local
  realism by two entangled n-dimensional systems are stronger than for two
  qubits},\ }\href@noop {} {\bibfield  {journal} {\bibinfo  {journal} {Physical
  Review Letters}\ }\textbf {\bibinfo {volume} {85}},\ \bibinfo {pages} {4418}
  (\bibinfo {year} {2000})}\BibitemShut {NoStop}%
\bibitem [{\citenamefont {Collins}\ \emph {et~al.}(2002)\citenamefont
  {Collins}, \citenamefont {Gisin}, \citenamefont {Linden}, \citenamefont
  {Massar},\ and\ \citenamefont {Popescu}}]{collins2002bell}%
  \BibitemOpen
  \bibfield  {author} {\bibinfo {author} {\bibfnamefont {D.}~\bibnamefont
  {Collins}}, \bibinfo {author} {\bibfnamefont {N.}~\bibnamefont {Gisin}},
  \bibinfo {author} {\bibfnamefont {N.}~\bibnamefont {Linden}}, \bibinfo
  {author} {\bibfnamefont {S.}~\bibnamefont {Massar}},\ and\ \bibinfo {author}
  {\bibfnamefont {S.}~\bibnamefont {Popescu}},\ }\bibfield  {title} {\bibinfo
  {title} {Bell inequalities for arbitrarily high-dimensional systems},\
  }\href@noop {} {\bibfield  {journal} {\bibinfo  {journal} {Physical review
  letters}\ }\textbf {\bibinfo {volume} {88}},\ \bibinfo {pages} {040404}
  (\bibinfo {year} {2002})}\BibitemShut {NoStop}%
\bibitem [{\citenamefont {Dada}\ \emph {et~al.}(2011)\citenamefont {Dada},
  \citenamefont {Leach}, \citenamefont {Buller}, \citenamefont {Padgett},\ and\
  \citenamefont {Andersson}}]{dada2011experimental}%
  \BibitemOpen
  \bibfield  {author} {\bibinfo {author} {\bibfnamefont {A.~C.}\ \bibnamefont
  {Dada}}, \bibinfo {author} {\bibfnamefont {J.}~\bibnamefont {Leach}},
  \bibinfo {author} {\bibfnamefont {G.~S.}\ \bibnamefont {Buller}}, \bibinfo
  {author} {\bibfnamefont {M.~J.}\ \bibnamefont {Padgett}},\ and\ \bibinfo
  {author} {\bibfnamefont {E.}~\bibnamefont {Andersson}},\ }\bibfield  {title}
  {\bibinfo {title} {Experimental high-dimensional two-photon entanglement and
  violations of generalized bell inequalities},\ }\href@noop {} {\bibfield
  {journal} {\bibinfo  {journal} {Nature Physics}\ }\textbf {\bibinfo {volume}
  {7}},\ \bibinfo {pages} {677} (\bibinfo {year} {2011})}\BibitemShut {NoStop}%
\bibitem [{\citenamefont {Campbell}\ \emph {et~al.}(2012)\citenamefont
  {Campbell}, \citenamefont {Anwar},\ and\ \citenamefont
  {Browne}}]{campbell2012magic}%
  \BibitemOpen
  \bibfield  {author} {\bibinfo {author} {\bibfnamefont {E.~T.}\ \bibnamefont
  {Campbell}}, \bibinfo {author} {\bibfnamefont {H.}~\bibnamefont {Anwar}},\
  and\ \bibinfo {author} {\bibfnamefont {D.~E.}\ \bibnamefont {Browne}},\
  }\bibfield  {title} {\bibinfo {title} {Magic-state distillation in all prime
  dimensions using quantum reed-muller codes},\ }\href@noop {} {\bibfield
  {journal} {\bibinfo  {journal} {Physical Review X}\ }\textbf {\bibinfo
  {volume} {2}},\ \bibinfo {pages} {041021} (\bibinfo {year}
  {2012})}\BibitemShut {NoStop}%
\bibitem [{\citenamefont {Zhang}\ \emph {et~al.}(2016)\citenamefont {Zhang},
  \citenamefont {Roux}, \citenamefont {Konrad}, \citenamefont {Agnew},
  \citenamefont {Leach},\ and\ \citenamefont {Forbes}}]{zhang2016engineering}%
  \BibitemOpen
  \bibfield  {author} {\bibinfo {author} {\bibfnamefont {Y.}~\bibnamefont
  {Zhang}}, \bibinfo {author} {\bibfnamefont {F.~S.}\ \bibnamefont {Roux}},
  \bibinfo {author} {\bibfnamefont {T.}~\bibnamefont {Konrad}}, \bibinfo
  {author} {\bibfnamefont {M.}~\bibnamefont {Agnew}}, \bibinfo {author}
  {\bibfnamefont {J.}~\bibnamefont {Leach}},\ and\ \bibinfo {author}
  {\bibfnamefont {A.}~\bibnamefont {Forbes}},\ }\bibfield  {title} {\bibinfo
  {title} {Engineering two-photon high-dimensional states through quantum
  interference},\ }\href@noop {} {\bibfield  {journal} {\bibinfo  {journal}
  {Science advances}\ }\textbf {\bibinfo {volume} {2}},\ \bibinfo {pages}
  {e1501165} (\bibinfo {year} {2016})}\BibitemShut {NoStop}%
\bibitem [{\citenamefont {Barreiro}\ \emph {et~al.}(2005)\citenamefont
  {Barreiro}, \citenamefont {Langford}, \citenamefont {Peters},\ and\
  \citenamefont {Kwiat}}]{barreiro2005generation}%
  \BibitemOpen
  \bibfield  {author} {\bibinfo {author} {\bibfnamefont {J.~T.}\ \bibnamefont
  {Barreiro}}, \bibinfo {author} {\bibfnamefont {N.~K.}\ \bibnamefont
  {Langford}}, \bibinfo {author} {\bibfnamefont {N.~A.}\ \bibnamefont
  {Peters}},\ and\ \bibinfo {author} {\bibfnamefont {P.~G.}\ \bibnamefont
  {Kwiat}},\ }\bibfield  {title} {\bibinfo {title} {Generation of
  hyperentangled photon pairs},\ }\href@noop {} {\bibfield  {journal} {\bibinfo
   {journal} {Physical review letters}\ }\textbf {\bibinfo {volume} {95}},\
  \bibinfo {pages} {260501} (\bibinfo {year} {2005})}\BibitemShut {NoStop}%
\bibitem [{\citenamefont {Graham}\ \emph {et~al.}(2015)\citenamefont {Graham},
  \citenamefont {Bernstein}, \citenamefont {Wei}, \citenamefont {Junge},\ and\
  \citenamefont {Kwiat}}]{graham2015superdense}%
  \BibitemOpen
  \bibfield  {author} {\bibinfo {author} {\bibfnamefont {T.~M.}\ \bibnamefont
  {Graham}}, \bibinfo {author} {\bibfnamefont {H.~J.}\ \bibnamefont
  {Bernstein}}, \bibinfo {author} {\bibfnamefont {T.-C.}\ \bibnamefont {Wei}},
  \bibinfo {author} {\bibfnamefont {M.}~\bibnamefont {Junge}},\ and\ \bibinfo
  {author} {\bibfnamefont {P.~G.}\ \bibnamefont {Kwiat}},\ }\bibfield  {title}
  {\bibinfo {title} {Superdense teleportation using hyperentangled photons},\
  }\href@noop {} {\bibfield  {journal} {\bibinfo  {journal} {Nature
  communications}\ }\textbf {\bibinfo {volume} {6}},\ \bibinfo {pages} {1}
  (\bibinfo {year} {2015})}\BibitemShut {NoStop}%
\bibitem [{\citenamefont {Rafsanjani}(2019)}]{rafsanjani2019sorting}%
  \BibitemOpen
  \bibfield  {author} {\bibinfo {author} {\bibfnamefont {S.~M.~H.}\
  \bibnamefont {Rafsanjani}},\ }\bibfield  {title} {\bibinfo {title}
  {Sorting-based approach to multiphoton interference},\ }\href@noop {}
  {\bibfield  {journal} {\bibinfo  {journal} {Optics letters}\ }\textbf
  {\bibinfo {volume} {44}},\ \bibinfo {pages} {4993} (\bibinfo {year}
  {2019})}\BibitemShut {NoStop}%
\bibitem [{\citenamefont {Torres}\ and\ \citenamefont
  {Torner}(2011)}]{torres2011twisted}%
  \BibitemOpen
  \bibfield  {author} {\bibinfo {author} {\bibfnamefont {J.~P.}\ \bibnamefont
  {Torres}}\ and\ \bibinfo {author} {\bibfnamefont {L.}~\bibnamefont
  {Torner}},\ }\href@noop {} {\emph {\bibinfo {title} {Twisted photons:
  applications of light with orbital angular momentum}}}\ (\bibinfo
  {publisher} {John Wiley \& Sons},\ \bibinfo {year} {2011})\BibitemShut
  {NoStop}%
\bibitem [{\citenamefont {Walborn}\ \emph {et~al.}(2010)\citenamefont
  {Walborn}, \citenamefont {Monken}, \citenamefont {P{\'a}dua},\ and\
  \citenamefont {Ribeiro}}]{walborn2010spatial}%
  \BibitemOpen
  \bibfield  {author} {\bibinfo {author} {\bibfnamefont {S.~P.}\ \bibnamefont
  {Walborn}}, \bibinfo {author} {\bibfnamefont {C.}~\bibnamefont {Monken}},
  \bibinfo {author} {\bibfnamefont {S.}~\bibnamefont {P{\'a}dua}},\ and\
  \bibinfo {author} {\bibfnamefont {P.~S.}\ \bibnamefont {Ribeiro}},\
  }\bibfield  {title} {\bibinfo {title} {Spatial correlations in parametric
  down-conversion},\ }\href@noop {} {\bibfield  {journal} {\bibinfo  {journal}
  {Physics Reports}\ }\textbf {\bibinfo {volume} {495}},\ \bibinfo {pages} {87}
  (\bibinfo {year} {2010})}\BibitemShut {NoStop}%
\bibitem [{\citenamefont {Weihs}\ and\ \citenamefont
  {Zeilinger}(2001)}]{weihs2001photon}%
  \BibitemOpen
  \bibfield  {author} {\bibinfo {author} {\bibfnamefont {G.}~\bibnamefont
  {Weihs}}\ and\ \bibinfo {author} {\bibfnamefont {A.}~\bibnamefont
  {Zeilinger}},\ }\bibfield  {title} {\bibinfo {title} {Photon statistics at
  beam-splitters: an essential tool in quantum information and teleportation},\
  }\href@noop {} {\bibfield  {journal} {\bibinfo  {journal} {Coherence and
  Statistics of Photons and Atoms}\ ,\ \bibinfo {pages} {262}} (\bibinfo {year}
  {2001})}\BibitemShut {NoStop}%
\bibitem [{\citenamefont {Ekert}(1991)}]{ekert1991quantum}%
  \BibitemOpen
  \bibfield  {author} {\bibinfo {author} {\bibfnamefont {A.~K.}\ \bibnamefont
  {Ekert}},\ }\bibfield  {title} {\bibinfo {title} {Quantum cryptography based
  on bell’s theorem},\ }\href@noop {} {\bibfield  {journal} {\bibinfo
  {journal} {Physical Review Letters}\ }\textbf {\bibinfo {volume} {67}},\
  \bibinfo {pages} {661} (\bibinfo {year} {1991})}\BibitemShut {NoStop}%
\end{thebibliography}

%

\end{document}